\documentclass[sigconf]{acmart}

\usepackage{multirow}
\usepackage{booktabs}
\usepackage{enumitem}
\usepackage[normalem]{ulem}
\usepackage{natbib}
\usepackage{pifont}
\usepackage{amsmath}
\usepackage{color, xspace}

\usepackage{color}
\usepackage{tikz}
\usepackage{xcolor}
\usepackage{xspace}
\usepackage{booktabs}
\usepackage{multirow}
\usepackage{makecell}
\usepackage[edges]{forest}

\AtBeginDocument{%
  \providecommand\BibTeX{{%
    \normalfont B\kern-0.5em{\scshape i\kern-0.25em b}\kern-0.8em\TeX}}}
\copyrightyear{2024} 
\acmYear{2024} 
\setcopyright{acmlicensed}\acmConference[KDD '24]{Proceedings of the 30th
ACM SIGKDD Conference on Knowledge Discovery and Data Mining}{August
25--29, 2024}{Barcelona, Spain}
\acmBooktitle{Proceedings of the 30th ACM SIGKDD Conference on Knowledge
Discovery and Data Mining (KDD '24), August 25--29, 2024, Barcelona, Spain}
\acmDOI{10.1145/3637528.3671571}
\acmISBN{979-8-4007-0490-1/24/08}

\begin{document}

\title{ERASE: Benchmarking Feature Selection Methods for Deep Recommender Systems}
\makeatletter
\def\@ACM@checkaffil{%
    \if@ACM@instpresent\else
    \ClassWarningNoLine{\@classname}{No institution present for an affiliation}%
    \fi
    \if@ACM@citypresent\else
    \ClassWarningNoLine{\@classname}{No city present for an affiliation}%
    \fi
    \if@ACM@countrypresent\else
        \ClassWarningNoLine{\@classname}{No country present for an affiliation}%
    \fi
}
\makeatother

\author{Pengyue Jia}
\authornote{Authors contributed equally to this research.}
\email{jia.pengyue@my.cityu.edu.hk}
\affiliation{%
  \institution{City University of Hong Kong}
}

\author{Yejing Wang}
\authornotemark[1]
\email{yejing.wang@my.cityu.edu.hk}
\affiliation{
    \institution{City University of Hong Kong}
}

\author{Zhaocheng Du}
\authornotemark[1]
\email{zhaochengdu@huawei.com}
\affiliation{
    \institution{Huawei Noah's Ark Lab}
}

\author{Xiangyu Zhao}
\authornote{Corresponding author.}
\email{xianzhao@cityu.edu.hk}
\affiliation{
    \institution{City University of Hong Kong}
}

\author{Yichao Wang}
\email{wangyichao5@huawei.com}
\affiliation{
    \institution{Huawei Noah's Ark Lab}
}

\author{Bo Chen}
\email{chenbo116@huawei.com}
\affiliation{
    \institution{Huawei Noah's Ark Lab}
}

\author{Wanyu Wang}
\email{wanyuwang4-c@my.cityu.edu.hk}
\affiliation{
    \institution{City University of Hong Kong}
}

\author{Huifeng Guo}
\authornotemark[2]
\email{huifeng.guo@huawei.com}
\affiliation{
    \institution{Huawei Noah's Ark Lab}
}

\author{Ruiming Tang}
\email{tangruiming@huawei.com}
\affiliation{
    \institution{Huawei Noah's Ark Lab}
}

\renewcommand{\shortauthors}{Pengyue Jia, et al.}

\begin{abstract}
Deep Recommender Systems (DRS) are increasingly dependent on a large number of feature fields for more precise recommendations. Effective feature selection methods are consequently becoming critical for further enhancing the accuracy and optimizing storage efficiencies to align with the deployment demands. This research area, particularly in the context of DRS, is nascent and faces three core challenges. Firstly, variant experimental setups across research papers often yield unfair comparisons, obscuring practical insights. Secondly, the existing literature's lack of detailed analysis on selection attributes, based on large-scale datasets and a thorough comparison among selection techniques and DRS backbones, restricts the generalizability of findings and impedes deployment on DRS.
Lastly, research often focuses on comparing the peak performance achievable by feature selection methods. This approach is typically computationally infeasible for identifying the optimal hyperparameters and overlooks evaluating the robustness and stability of these methods. To bridge these gaps, this paper presents ERASE, a comprehensive b\textbf{E}nchma\textbf{R}k for fe\textbf{A}ture \textbf{SE}lection for DRS. ERASE comprises a thorough evaluation of eleven feature selection methods, covering both traditional and deep learning approaches, across four public datasets, private industrial datasets, and a real-world commercial platform, achieving significant enhancement.
Our code is available online\footnote{\url{https://github.com/Applied-Machine-Learning-Lab/ERASE}} for ease of reproduction.

\end{abstract}

\ccsdesc[500]{Information systems~Recommender systems}
\keywords{Benchmark, Feature Selection, Deep Recommender System}

\maketitle
\section{Introduction}

Recommender systems have become indispensable in many sectors ranging from e-commence to content streaming in the realm of information explosion~\cite{rendle2009bpr,rendle2010fm}. With the application of deep learning techniques, Deep Recommender System (DRS) exhibits amplified prediction ability of user preference, providing personalized experience and dominating the deployment landscape~\cite{cheng2016wide,wang2017deep,guo2017deepfm}.

To improve the accuracy of recommendations, DRS is progressively integrating an expanding array of feature fields into their predictive models, which can number in the hundreds or even thousands~\cite{heaton2016empirical}. The significance of each feature varies, resulting in the accumulation of superfluous or extraneous features. Consequently, feature selection, which concentrates on pinpointing and leveraging the most critical features, is becoming increasingly crucial in modern DRS~\cite{zheng2023automl,chen2024comprehensive}. One immediate benefit of feature selection is the enhancement of prediction performance, achieved by eliminating non-contributory features that could otherwise adversely affect predictions. From an industrial viewpoint, selecting predictive features is also essential for meeting deployment criteria regarding memory usage since unnecessary storage demands can often be inflated by the presence of redundant features.

Hand-crafted feature selection usually requires lots of expert knowledge and labor efforts, which is usually infeasible to achieve optimal results when the candidate set contains thousands of features. Researchers have designed various methods to automatically select predictive features, including statistical methods~\cite{tibshirani1996lasso,liu1995chi2,vergara2014review}, learning methods~\cite{chen2016xgboost,friedman2001greedy,breiman2001rf}, and agent-based methods~\cite{liu2019automating,fan2020autofs,zhao2020simplifying}. Despite the satisfactory results these methods achieved, they also reveal the following issues hindering the development of this field:
\begin{itemize}[leftmargin=*]
    \item \textbf{Experimental Differences.} Recent years have witnessed a variety of methods designed for DRS~\cite{wang2022autofield,lin2022adafs,guo2022lpfs,lee2023mvfs,lyu2023optfs,Fperm,wang2023sfs}. However, these works conduct experiments with varied settings. For example, MultiSFS~\cite{wang2023sfs} is designed to select features for multi-task DRS. SHARK~\cite{Fperm} suggests the feature selection method F-permutation and a quantization method. Optfs~\cite{lyu2023optfs} selects the feature from the value-level while others mainly select from the field-level. These experimental differences usually lead to unfair or unavailable comparisons, making the subsequent researchers struggle to generate practical insights. 
    \item \textbf{Insufficiency.}
    Existing feature selection benchmarks are predominantly tailored for conventional downstream tasks, such as classification~\cite{bolon2014review,bommert2020benchmark}. They are primarily built upon synthetic or domain-specific datasets~\cite{bolon2013review,forman2003extensive,darshan2018performance}, which diverges significantly from the complex, large-scale datasets encountered in DRS. This divergence results in a notable deficiency in guiding feature selection specifically for DRS, due to the benchmarks' disparate scope and focus.
    DeepLasso~\cite{deeplasso}, while being among the benchmarks most closely related to DRS, primarily addresses tabular learning—a context that, despite similarities, does not fully align with the intricacies of recommendation tasks. Furthermore, such benchmarks often limit their exploration to traditional selection methods and rely on datasets of a much smaller scale, thereby omitting crucial, in-depth analysis from an industrial perspective for DRS.
    Moreover, while related literature surveys~\cite{chen2024comprehensive,zheng2023automl} provide comprehensive reviews of the field, they fall short in offering empirical evidence or experimental results that could motivate practical deployments in DRS, lacking the necessary data-driven support to inform and inspire future research directions.
    \item \textbf{Assessment Deficit.} 
    A pivotal hyperparameter for feature selection methods is the number of features to be selected, denoted as $k$ in this study. Feature selection methods often exhibit significant variability in performance across different values of $k$, and the optimal $k$ is not the same for all methods~\cite{wang2022autofield,lin2022adafs,lyu2023optfs}. This variability on $k$ introduces two significant challenges in evaluating feature selection methods.
    First, the direct comparison of methods at their respective optimal $k$ may not constitute a fair assessment. Such comparisons fail to account for the different memory requirements associated with varying optimal $k$ values for different selection methods. Furthermore, this approach neglects the performance variability under sub-optimal hyperparameter settings, thereby obscuring insights into the methods' robustness and stability across a range of $k$ values.
    Second, the exhaustive search for the optimal $k$ across the entire spectrum of possible values is time-consuming and computationally intensive. This necessitates an evaluation methodology capable of effectively assessing a method's performance based on partial $k$ values, offering a more efficient means to gauge feature selection effectiveness without finishing complete iterations.
\end{itemize}

To address these issues, we propose ERASE, a comprehensive b\textbf{E}nchma\textbf{R}k for fe\textbf{A}ture \textbf{S}election for DRS. 
ERASE initiates a unified and fair experimental framework, minimizing experimental discrepancies across various selection methods.
Remarkably, ERASE pioneers as the first feature selection benchmark with a focus on DRS tasks, incorporating both prevalent DRS feature selection techniques and conventional methods. It introduces a novel taxonomy to classify these methods and unearth intrinsic patterns among groups of methods. By evaluating the performance on widely used public datasets and authentic industrial production datasets—through both offline comparison and online testing—ERASE furnishes strong empirical support, facilitating the generation of actionable insights.
In its endeavor to provide a comprehensive assessment of feature selection methods, ERASE contrasts the optimal performance of compared selection methods alongside their outcomes under specific deployment prerequisites. Additionally, we introduce a novel metric, AUKC, specifically crafted for assessing the robustness and stability of feature selection methods across a range of feature quantities $k$, thereby addressing the critical lack of such evaluative metrics.
We summarize our major contributions as follows:

\begin{itemize}[leftmargin=*]
    \item We present ERASE, a comprehensive benchmark for DRS feature selection methods, providing a fair comparison for emerging selection techniques with various datasets and DRS backbones. To the best of our knowledge, we are the first to focus on benchmarking feature selection methods for recommendation tasks. %
    \item We recognize the assessment shortfall linked to the substantial dependency of selection efficacy on the hyperparameter $k$, and in response, we introduce a novel evaluation metric, AUKC. This metric is designed to evaluate the robustness and stability of feature selection methods, bridging the existing gap in assessment.
    \item We carry out thorough experiments across four widely used public datasets and real-world industrial production datasets, yielding insights from various angles. Notably, our experimental findings have guided optimizations in our online platform, achieving a 20\% reduction in latency without compromising effectiveness, validating the practical utility of our benchmark.
\end{itemize}

\section{Becnchmark Design}

\definecolor{mygreen}{HTML}{9ac7bf}
\definecolor{myblue}{HTML}{a9c4eb}
\definecolor{myred}{HTML}{ea6b66}
\definecolor{myyellow}{HTML}{ffe599}

\tikzstyle{my-box}=[
    rectangle,
    draw=white,
    rounded corners,
    align=left,
    text opacity=1,
    minimum height=1.5em,
    minimum width=5em,
    inner sep=2pt,
    fill opacity=.8,
    line width=0.8pt,
]

\tikzstyle{leaf-head}=[my-box, minimum height=1.5em,
    draw=gray!80, %
    fill=gray!15,  %
    text=black, font=\normalsize,
    inner xsep=2pt,
    inner ysep=4pt,
    line width=0.8pt,
]

\tikzstyle{leaf-dataset}=[my-box, minimum height=1.5em,
    draw=mygreen, %
    fill=mygreen,  %
    text=black, font=\normalsize,
    inner xsep=2pt,
    inner ysep=4pt,
    line width=0.8pt,
]

\tikzstyle{leaf-fs}=[my-box, minimum height=1.5em,
    draw=myblue, %
    fill=myblue,  %
    text=black, font=\normalsize,
    inner xsep=2pt,
    inner ysep=4pt,
    line width=0.8pt,
]
\tikzstyle{leaf-bbm}=[my-box, minimum height=1.5em,
    draw=myred, %
    fill=myred,  %
    text=black, font=\normalsize,
    inner xsep=2pt,
    inner ysep=4pt,
    line width=0.8pt,
]
\tikzstyle{leaf-metrics}=[my-box, minimum height=1.5em,
    draw=myyellow, %
    fill=myyellow,  %
    text=black, font=\normalsize,
    inner xsep=2pt,
    inner ysep=4pt,
    line width=0.8pt,
]

\tikzstyle{modelnode-dataset}=[my-box, minimum height=1.5em,
    draw=mygreen!70, %
    fill=mygreen!70,  %
    text=black, font=\normalsize,
    inner xsep=2pt,
    inner ysep=4pt,
    line width=0.8pt,
]

\tikzstyle{modelnode-fs}=[my-box, minimum height=1.5em,
    draw=myblue!70, %
    fill=myblue!70,  %
    text=black, font=\normalsize,
    inner xsep=2pt,
    inner ysep=4pt,
    line width=0.8pt,
]
\tikzstyle{modelnode-bbm}=[my-box, minimum height=1.5em,
    draw=myred!70, %
    fill=myred!70,  %
    text=black, font=\normalsize,
    inner xsep=2pt,
    inner ysep=4pt,
    line width=0.8pt,
]
\tikzstyle{modelnode-metrics}=[my-box, minimum height=1.5em,
    draw=myyellow!70, %
    fill=myyellow!70,  %
    text=black, font=\normalsize,
    inner xsep=2pt,
    inner ysep=4pt,
    line width=0.8pt,
]

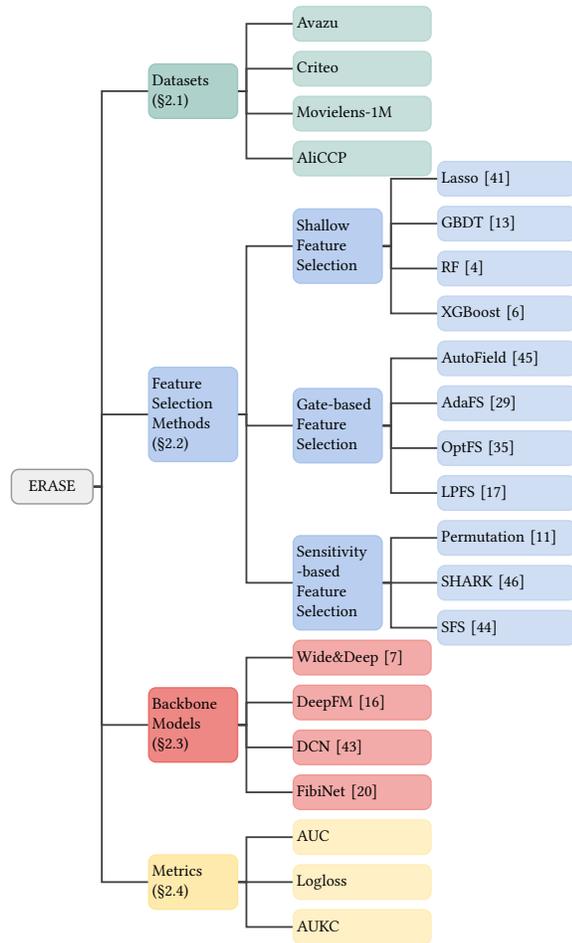
\begin{figure}[]
    \centering
    \resizebox{0.9\linewidth}{!}{
        \begin{forest}
            forked edges,
            for tree={
                grow=east,
                reversed=true,
                anchor=base west,
                parent anchor=east,
                child anchor=west,
                base=left,
                font=\normalsize,
                rectangle,
                draw=white,
                rounded corners,
                align=left,
                minimum width=1em,
                edge+={darkgray, line width=1pt},
                s sep=5pt,
                l sep=30pt,
                inner xsep=3pt,
                inner ysep=3pt,
                line width=1pt,
                ver/.style={rotate=90, child anchor=north, parent anchor=south, anchor=center},
            }, 
            [%
                ERASE,leaf-head
                [
                     Datasets \\ (\S\ref{sec:datasets}), leaf-dataset,text width=5em
                     [
                        \href{https://www.kaggle.com/competitions/avazu-ctr-prediction}{Avazu}, modelnode-dataset, text width=8em
                     ]
                     [
                        \href{https://ailab.criteo.com/ressources/}{Criteo}, modelnode-dataset, text width=8em
                     ]
                     [
                        \href{https://grouplens.org/datasets/movielens/1m/}{Movielens-1M}, modelnode-dataset, text width=8em
                     ]
                     [
                        \href{https://tianchi.aliyun.com/dataset/408}{AliCCP}, modelnode-dataset, text width=8em
                     ]
                ]
                [
                    Feature \\ Selection \\ Methods \\ (\S\ref{sec:fs}), leaf-fs,text width=5em
                    [
                        Shallow \\ Feature \\ Selection, leaf-fs, text width=5em
                        [Lasso~\cite{tibshirani1996lasso}, modelnode-fs, text width=8em]
                        [GBDT~\cite{friedman2001greedy}, modelnode-fs, text width=8em]
                        [RF~\cite{breiman2001rf}, modelnode-fs, text width=8em]
                        [XGBoost~\cite{chen2016xgboost}, modelnode-fs, text width=8em]
                    ]
                    [
                        Gate-based \\ Feature \\ Selection, leaf-fs, text width=5em
                        [AutoField~\cite{wang2022autofield}, modelnode-fs, text width=8em]
                        [AdaFS~\cite{lin2022adafs}, modelnode-fs, text width=8em]
                        [OptFS~\cite{lyu2023optfs}, modelnode-fs, text width=8em]
                        [LPFS~\cite{guo2022lpfs},modelnode-fs, text width=8em]
                    ]
                    [
                        Sensitivity \\ -based \\ Feature \\ Selection, leaf-fs, text width=5em
                        [Permutation~\cite{fisher2019allpermutation}, modelnode-fs, text width=8em]
                        [SHARK~\cite{Fperm}, modelnode-fs, text width=8em]
                        [SFS~\cite{wang2023sfs}, modelnode-fs, text width=8em]
                    ]
                ]
                [
                    Backbone \\ Models \\ (\S\ref{sec:backbone_models}), leaf-bbm,text width=5em
                    [ 
                        Wide\&Deep~\cite{cheng2016wide}, modelnode-bbm, text width=8em
                    ]
                    [ DeepFM~\cite{guo2017deepfm},modelnode-bbm, text width=8em]
                    [ DCN~\cite{wang2017deep},modelnode-bbm, text width=8em]
                    [ FibiNet~\cite{huang2019fibinet},modelnode-bbm, text width=8em]
                ]
                [
                    Metrics \\ (\S\ref{sec:metrics}), leaf-metrics,text width=5em
                    [AUC, modelnode-metrics,text width=8em]
                    [Logloss, modelnode-metrics,text width=8em]
                    [AUKC, modelnode-metrics,text width=8em]
                ]
            ]
        \end{forest}
    }
    \caption{Benchmark Overview.}
    \label{fig:overview}
    \vspace{-4mm}
\end{figure}

In this section, we will give an overview of our benchmark. As shown in Figure~\ref{fig:overview}, the benchmark consists of four components: dataset, feature selection methods, backbone models, and metrics. %

\subsection{Datasets} \label{sec:datasets}

To comprehensively evaluate the effectiveness of different feature selection methods, we select four public datasets for our experiments. \uline{\textbf{Avazu}}\footnote{https://www.kaggle.com/competitions/avazu-ctr-prediction} and \uline{\textbf{Criteo}}\footnote{https://ailab.criteo.com/ressources/} are selected because they are frequently used dataset for studying feature selection in DRS~\cite{wang2022autofield,lin2022adafs,lyu2023optfs}. To compare the performance of different methods with a small feature set, we choose the popular \uline{\textbf{Movielens-1M}}\footnote{https://grouplens.org/datasets/movielens/1m/} dataset in the recommendation field. Additionally, to compare the performance of different methods in scenarios closer to real-world recommender systems, we supplement with the \uline{\textbf{AliCCP}}\footnote{https://tianchi.aliyun.com/dataset/408} dataset, which possesses user and item features and includes a rich set of 85,316,519 interaction samples. The statistics of datasets and the detailed introduction are illustrated in Appendix~\ref{appendix:datasets}.

\subsection{Feature Selection Methods} \label{sec:fs}

\begin{figure*}
    \centering
    \includegraphics[width=0.85\textwidth]{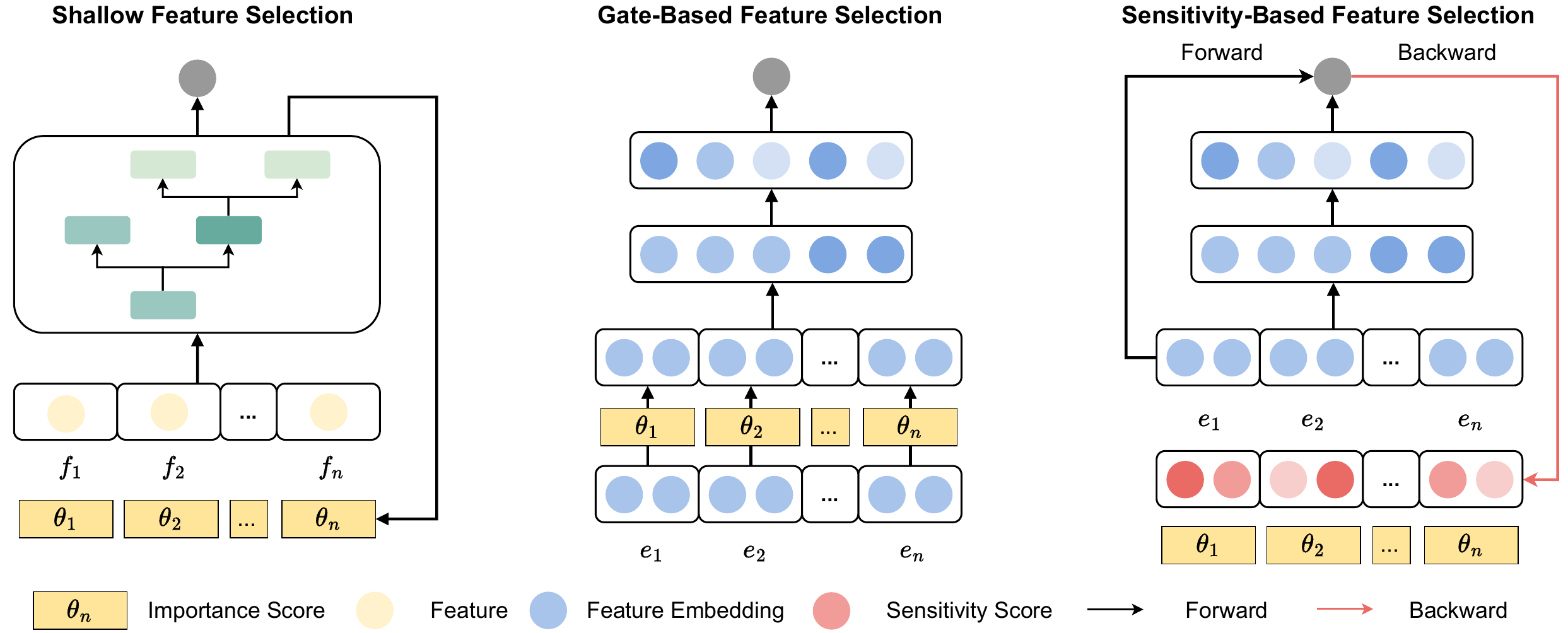}
    \caption{Overview of three categories of feature selection methods in deep recommender systems. Shallow methods typically use statistical algorithms to assign feature importance to each field. Gate-based methods, on the other hand, assign gates to feature embeddings and consider the gate values as the importance of the corresponding features. Sensitivity-based methods derive parameter sensitivity from backward propagation steps and calculate feature importance accordingly.}
    \label{fig:3-category}
\end{figure*}

We list all feature selection methods in this work and their attributes in Table~\ref{tab:methods}. There are three feasible dimensions to classify these methods. 1) \textbf{Training strategy.} Based on the training strategy, the methods can be single-stage or two-stage. Single-stage usually directly integrates the feature selection module into the original model without changing the training logic. In contrast, two-stage methods contain searching and retraining phases. Informative feature fields or feature values are selected in the searching phase, and the backbone model is trained with these selected features in the retraining phase. 2) \textbf{Selection type.} There are two types of selection in feature selection methods: soft selection and hard selection. The soft selection offers a mask to affect inputs. Features multiplied with 0 are considered filtered out. For the hard selection, features are removed directly from the inputs. 3) \textbf{Selection technique.} Depending on the techniques used for selection, as shown in Figure~\ref{fig:3-category}, we divide methods contained in our benchmark into three categories: shallow feature selection, gate-based selection, and sensitivity-based selection. %

Due to the unbalanced distribution of feature selection methods classifying on training strategy and selection type, we elaborate our work based on selection technique classification. %
Specifically, our benchmark contains the following methods:

\noindent\textbf{1. Shallow Feature Selection}

\begin{itemize}[leftmargin=*]
    \item \textbf{Lasso~\cite{tibshirani1996lasso}.} The least absolute shrinkage and selection operator (Lasso) algorithm is a traditional and useful method in machine learning. It performs both variable selection and regularization to improve the model performance.
    \item \textbf{GBDT~\cite{friedman2001greedy}.} Gradient-boosted decision tree (GBDT) achieves superior performance by continually adding trees to fit residuals. By aggregating the feature importance scores from each tree, it also serves as an effective method for feature selection.
    \item \textbf{RandomForest~\cite{breiman2001rf} (RF).} RandomForest derives the feature importance by measuring how much each feature decreases the impurity in a tree, commonly using Gini impurity or entropy.
    \item \textbf{XGBoost \cite{chen2016xgboost}.} Extreme Gradient Boosting (XGBoost) ranks the importance of features by calculating the improvement of every feature on the final performance.
\end{itemize}

\begin{table}[t]
\centering
\caption{Methods Overview. For the type column, "Shallow" represents shallow feature selection methods, "Gate" represents gate-based feature selection methods, and "Sensitivity" represents the sensitivity-based feature selection methods. For the single-stage, two-stage, soft selection, and hard selection columns, \ding{52} represents applicable, \ding{56} represents not applicable, and \ding{52}$\star$ represents that this method can be applicable after appropriate modifications.}
\vspace{-3mm}
\label{tab:methods}
\resizebox{\linewidth}{!}{
\begin{tabular}{cccccc} 
\toprule
Methods     & Type              & Single-stage & Two-stage & Soft Selection & Hard Selection  \\ 
\midrule
Lasso       & Shallow           &    \ding{56}          &    \ding{52}       &     \ding{56}          &       \ding{52}          \\
GBDT        & Shallow           &    \ding{56}          &    \ding{52}       &     \ding{56}          &       \ding{52}          \\
RF          & Shallow           &    \ding{56}          &    \ding{52}       &     \ding{56}          &       \ding{52}          \\
XGBoost     & Shallow           &    \ding{56}          &    \ding{52}       &     \ding{56}          &       \ding{52}          \\
AutoField   & Gate        &    \ding{56}          &    \ding{52}       &      \ding{56}         &       \ding{52}           \\
AdaFS       & Gate        &    \ding{52}          &    \ding{56}       &      \ding{52}         &       \ding{56}          \\
OptFS       & Gate        &    \ding{56}          &    \ding{52}       &      \ding{52}         &       \ \ \ding{52}$\star$          \\
LPFS        & Gate        &    \ding{52}          &    \ \ \ding{52}$\star$      &     \ding{52}          &      \ \ \ding{52}$\star$           \\
Permutation & Sensitivity &    \ding{56}          &    \ding{52}       &      \ding{56}         &       \ding{52}          \\
SHARK       & Sensitivity &    \ding{56}          &     \ding{52}      &      \ding{56}         &        \ding{52}         \\
SFS         & Sensitivity &    \ding{56}          &     \ding{52}      &      \ding{56}         &        \ding{52}         \\
\bottomrule
\end{tabular}}
\vspace{-3mm}
\end{table}

\noindent\textbf{2. Gate-based Feature Selection}
\begin{itemize}[leftmargin=*]
    \item \textbf{AutoField~\cite{wang2022autofield}.} AutoField is the first gate-based feature selection method in DRS. It designs a novel controller network to generate a 2-dimensional vector determining whether to choose this feature field. It is a two-stage method and selects features on the field level.
    \item \textbf{AdaFS~\cite{lin2022adafs}.} AdaFS is an algorithm that can generate a gate for each feature field adaptively. It is a single-stage method and selects features on the field level.
    \item \textbf{OptFS~\cite{lyu2023optfs}.} OptFS focuses on the feature value level selection. It trains a scalar for each feature value and continuously strengthens the constraints on sparsity as training progresses.
    \item \textbf{LPFS~\cite{guo2022lpfs}.} LPFS argues that the conclusion that features with smaller weights are less important than those with larger weights may not be correct. It proposes a kind of smoothed-$l^0$ function that can effectively select informative features. It is a single-stage method and focuses on the field level.
\end{itemize}

\noindent\textbf{3. Sensitivity-based Feature Selection}
\begin{itemize}[leftmargin=*]
    \item \textbf{Permutation~\cite{fisher2019allpermutation}.} Permutation works by randomly permuting the feature values at a time and measuring how much this permutation affects the performance of the model.
    \item \textbf{SHARK~\cite{Fperm}.} Shark takes the novel first-order component of Taylor expansion as the feature importance score for model prediction. It then prunes those features with lower scores from the embedding table to improve model performance and efficiency.
    \item \textbf{SFS~\cite{wang2023sfs}.} SFS takes the gradients of the gate for each feature field as the feature importance score. It is a two-stage method and selects on the feature field level.
\end{itemize}

\subsection{Backbone Models} \label{sec:backbone_models}

To comprehensively evaluate the effectiveness of feature selection methods, we choose four popular DRS models as backbones: 1) \uline{\textbf{Wide\&Deep}}: A classical model contains shallow and deep networks to capture feature interactions. 2) \uline{\textbf{DeepFM}}: A model with FM module to automatically learn feature interactions. 3) \uline{\textbf{DCN}}: A model with cross layers to study feature interactions. 4) \uline{\textbf{FibiNet}}: A model equipped with SENet and bilinear layer to adaptively learn feature importance and capture high-order feature interactions. The detailed introduction of these models is in Appendix~\ref{appendix:bbm}.

\subsection{Metrics} \label{sec:metrics}

In this paper, we focus on the Click-Through Rate (CTR) prediction task, so we take AUC and Logloss as the two main metrics in our benchmark. The detailed introduction is in Appendix~\ref{appendix:metrics}. Additionally, since there is currently no metric that comprehensively evaluates the performance of feature selection methods across varying numbers of selected features, we propose a new evaluation metric named ``AUKC'' to address this challenge.

\noindent\textbf{AUKC.} Current metrics cannot assess the robustness and stability of a specific feature selection method across different numbers of selections. Therefore, we introduce a new metric Area Under the $K$-performance Curve (AUKC) to fill this gap. AUKC measures the area under the performance curve of feature selection methods with different $K$. Specifically, AUKC is formalized as follows:
\begin{equation}
    AUKC = \frac{1}{|K|}\sum_{k=1}^{|K|}(AUC_{k}+AUC_{k-1}-1) \label{eq:AUKCCAl}
\end{equation}
where $|K|$ denotes the number of all feature fields, $AUC_k$ is the AUC score when selecting $k$ feature fields follow a specific feature selecting method. If there are no input features, the model would make random predictions so that $AUC_0 = 0.5$. 

In practice, due to resource constraints, it is generally not feasible to conduct experiments with every possible number of selections. Therefore, we further propose a more general form of AUKC to accommodate this change:
\begin{equation}
    AUKC=\frac{1}{|K|}\sum_{n=1}^{|N|}(AUC_{n,l}+AUC_{n,r}-1) \times \Delta l_{n}
\end{equation}
where $|K|$ denotes the number of all feature fields and $|N|$ is the number of segments across the entire length of the feature fields set. $AUC_{n,l}$ is the AUC corresponding to the number of features selected at the left endpoint of the $n$-th segment, and $AUC_{n,r}$ denotes the AUC with the number of selections at the right endpoint of the $n$th segment. $\Delta l_n$ represents the length of the $n$-th segment. The detailed process of the formula derivation is in Appendix~\ref{AUKC}.

AUKC and AUC share a similar conceptual basis, representing the area under a curve, and both have a value range from 0 to 1. However, unlike AUC, the endpoint of AUKC does not necessarily equal 1, and the values in the middle of the curve can exceed the endpoint value. This occurs because eliminating features with less information can result in a model that performs better than one using all available features. The AUKC metric considers the effectiveness of feature selection at different numbers of selected features, $k$, thereby providing a more comprehensive reflection of the efficacy of feature selection methods.
\section{Experiments}

\begin{table*}[ht]
\centering
\caption{Overall experimental results of feature selection for DeepFM and Wide\&Deep backbones.}
\label{overall_results}
\resizebox{0.8\textwidth}{!}{
\begin{tabular}{cccccccccc} 
\toprule
\multirow{2}{*}{Backbone model} & \multirow{2}{*}{Methods} & \multicolumn{2}{c}{Avazu}           & \multicolumn{2}{c}{Criteo}          & \multicolumn{2}{c}{Movielens-1M}    & \multicolumn{2}{c}{AliCCP}           \\ 
\cmidrule[\heavyrulewidth]{3-10}
                                &                          & AUC              & Logloss          & AUC              & Logloss          & AUC              & Logloss          & AUC              & Logloss           \\ 
\toprule
\multirow{12}{*}{DeepFM}        & no\_selection            & 0.78576          & 0.37680          & 0.80024          & \textbf{0.45321} & 0.79027          & 0.54124          & 0.61813          & 0.16176           \\
                                & Lasso                    & 0.78587          & 0.37654          & 0.79832          & 0.45483          & \uline{0.80683}  & \uline{0.52498}  & 0.61887          & 0.16174           \\
                                & GBDT                     & 0.76930          & 0.38571          & 0.80011          & 0.45329          & 0.78942          & 0.54234          & 0.61864          & \textbf{0.16150}  \\
                                & RF                       & \uline{0.78642}  & 0.37630          & 0.79920          & 0.45427          & 0.78920          & 0.54217          & 0.61772          & 0.16172           \\
                                & XGBoost                  & 0.76953          & 0.38548          & 0.80022          & 0.45333          & 0.78974          & 0.54190          & 0.61822          & 0.16172           \\
                                & AutoField                & 0.78611          & 0.37641          & 0.80022          & \uline{0.45325}  & \textbf{0.80722} & \textbf{0.52471} & \uline{0.61894}  & 0.16174           \\
                                & AdaFS                    & 0.78278          & 0.37922          & 0.79950          & 0.45405          & 0.78590          & 0.54676          & 0.61726          & \uline{0.16158}   \\
                                & OptFS                    & 0.78624          & \uline{0.37627}  & 0.79934          & 0.45505          & 0.79027          & 0.54124          & 0.61551          & 0.16195           \\
                                & LPFS                     & \textbf{0.78840} & \textbf{0.37601} & \textbf{0.80080} & 0.45364          & 0.78885          & 0.54276          & \textbf{0.61941} & 0.17864           \\
                                & Permutation              & 0.78624          & 0.37640          & 0.80006          & 0.45349          & 0.78974          & 0.54207          & 0.61844          & 0.16174           \\
                                & SHARK                    & 0.78611          & 0.37643          & \uline{0.80035}  & 0.45329          & 0.78888          & 0.54342          & 0.61840          & 0.16174           \\
                                & SFS                      & 0.78626          & 0.37631          & 0.80020          & 0.45330          & 0.78982          & 0.54194          & 0.61835          & 0.16168           \\ 
\midrule
\multirow{12}{*}{Wide\&Deep}      & no\_selection            & 0.78574          & 0.37679          & 0.79971          & 0.45346          & 0.79041          & 0.54163          & 0.62137          & 0.16154           \\
                                & Lasso                    & 0.78578          & 0.37665          & 0.79761          & 0.45526          & \textbf{0.80750} & \textbf{0.52400} & 0.62096          & 0.16159           \\
                                & GBDT                     & 0.76900          & 0.38583          & 0.79979          & 0.45337          & 0.78926          & 0.54252          & \uline{0.62187}  & 0.16146           \\
                                & RF                       & 0.78571          & 0.37664          & 0.79907          & 0.45400          & 0.78992          & 0.54170          & \textbf{0.62202} & 0.16132           \\
                                & XGBoost                  & 0.76886          & 0.38587          & \uline{0.80023}  & 0.45320          & 0.78893          & 0.54266          & 0.62174          & 0.16142           \\
                                & AutoField                & 0.78601          & 0.37653          & 0.79971          & 0.45348          & \uline{0.80736}  & \uline{0.52496}  & 0.62176          & 0.16134           \\
                                & AdaFS                    & \textbf{0.78607} & 0.37665          & 0.79952          & 0.45394          & 0.78703          & 0.54467          & 0.62106          & \uline{0.16123}   \\
                                & OptFS                    & 0.78582          & 0.37654          & \textbf{0.80111} & \textbf{0.45284} & 0.79041          & 0.54163          & 0.62053          & 0.16144           \\
                                & LPFS                     & 0.78606          & 0.37680          & 0.79953          & 0.45477          & 0.79072          & 0.54136          & 0.61916          & 0.17177           \\
                                & Permutation              & \textbf{0.78607} & \uline{0.37647}  & 0.79988          & 0.45342          & 0.78936          & 0.54226          & 0.62156          & 0.16138           \\
                                & SHARK                    & 0.78592          & 0.37655          & 0.80019          & \uline{0.45313}  & 0.78980          & 0.54146          & 0.62148          & 0.16154           \\
                                & SFS                      & 0.78592          & \textbf{0.37646} & 0.80010          & 0.45326          & 0.78993          & 0.54139          & 0.62165          & \textbf{0.16122}  \\ 
\bottomrule
\end{tabular}}
\end{table*}

In this section, we will provide extensive experimental results to answer following research questions:

\begin{itemize}[leftmargin=*]
    \item \textbf{RQ1:} In a fair and unified environment, how do different feature selection methods perform?
    \item\textbf{RQ2:} How do feature selection methods perform in terms of robustness and stability across various numbers of selections? %
\item \textbf{RQ3:} How is the efficiency of feature selection methods? %

\item\textbf{RQ4:} Do the performances of feature selection methods align between large-scale industrial datasets and public datasets?

\item\textbf{RQ5:} Do the performances of feature selection methods remain consistent online and offline?
\item \textbf{RQ6:} What is the relationship between feature rankings of different feature selection methods?
\end{itemize}

\subsection{Experimental Details}

We implement the benchmark framework based on Pytorch 1.11. In the training phase, we utilize the Adam optimizer with $\beta_1=0.9$, $\beta_2=0.999$, and $\epsilon=1\times10^{-8}$. We set the learning rate as 0.001, the batch size as 4,096, and the embedding size as 8. For the activation function, if the original papers do not emphasize a specific one, we use ReLU as the activation function. We release the repository of our benchmark online\footnote{\url{https://github.com/Applied-Machine-Learning-Lab/ERASE}}. 
We conduct each experimental setting three times and record its average metrics to mitigate the impact of experimental fluctuations.

\begin{figure*}[th]
    \centering
    \includegraphics[width=0.9\textwidth]{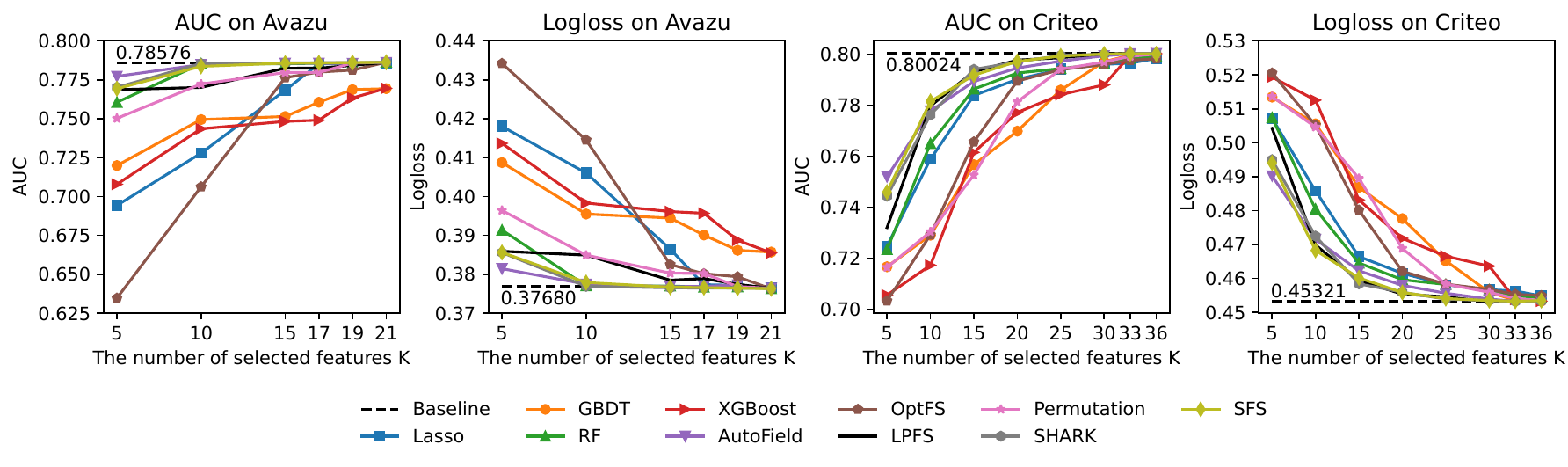}
    \caption{Experimental results of feature selection methods with DeepFM backbone, evaluated by AUC and Logloss.}
    \label{fig:k}
\end{figure*}

\subsection{Overall Performance (RQ1)}

\begin{table*}[t]
\centering
\caption{Experimental results of AUKC on Avazu and Criteo.}
\label{overall_results_AUKC}
\resizebox{0.9\textwidth}{!}{
\begin{tabular}{cccccccccccc} 
\toprule
\multicolumn{2}{c}{Methods}        & Lasso   & GBDT    & RF      & XGBoost & AutoField        & OptFS   & LPFS    & Permutation & SHARK            & SFS               \\ 
\midrule
\multirow{4}{*}{Avazu}  & DeepFM   & 0.43896 & 0.44430 & 0.49814 & 0.43264 & \textbf{0.50526} & 0.40455 & 0.49337 & 0.48526     & \uline{0.50267}  & 0.50146           \\
                        & WideDeep & 0.43845 & 0.44392 & 0.49780 & 0.43271 & \textbf{0.50537} & 0.40464 & 0.49349 & 0.48521     & \uline{0.50254}  & 0.50120           \\
                        & DCN      & 0.43909 & 0.44419 & 0.49920 & 0.43311 & \textbf{0.50658} & 0.40534 & 0.49472 & 0.48673     & \uline{0.50363}  & 0.50223           \\
                        & FibiNet  & 0.44410 & 0.44626 & 0.50475 & 0.43544 & \uline{0.50877}  & 0.40974 & 0.49866 & 0.49246     & \textbf{0.50918} & 0.50691           \\ 
\midrule
\multirow{4}{*}{Criteo} & DeepFM   & 0.52193 & 0.49894 & 0.52498 & 0.49388 & 0.53862          & 0.50410 & 0.53634 & 0.50311     & \uline{0.53886}  & \textbf{0.53998}  \\
                        & WideDeep & 0.52142 & 0.49821 & 0.52473 & 0.49338 & 0.53826          & 0.50369 & 0.53598 & 0.50279     & \uline{0.53857}  & \textbf{0.53972}  \\
                        & DCN      & 0.52245 & 0.49898 & 0.52582 & 0.49410 & 0.53926          & 0.50449 & 0.53690 & 0.50374     & \uline{0.53952}  & \textbf{0.54077}  \\
                        & FibiNet  & 0.52822 & 0.50459 & 0.53155 & 0.50006 & 0.54528          & 0.51048 & 0.54329 & 0.50964     & \uline{0.54580}  & \textbf{0.54693}  \\
\bottomrule
\end{tabular}}
\end{table*}

In this subsection, we compare the performance of feature selection methods with different backbone models. The complete experimental results for all four backbone models are illustrated in Appendix~\ref{appendix:overall}. For the two-stage approaches (Lasso, GBDT, RF, XGBoost, AutoField, Permutation, SHARK, and SFS), we experiment with different values of selected features during the retraining stage and adopt the best results. In the case of the single-stage approach, if it can also be modified to a two-stage method (OptFS and LPFS), we conduct experiments on both ways and record the optimal results; if it only supports single-stage (i.e., AdaFS), we directly document its results. From Table~\ref{overall_results}, we can observe the following points: 
\begin{itemize}[leftmargin=*]
    \item In general, gate-based methods show better performance compared to shallow feature selection and sensitivity-based feature selection. The likely reason is that compared to using traditional machine learning methods or solely relying on gradient information, leveraging the powerful expression and learning capabilities of deep neural networks allows for a more thorough exploration of feature importance.
    \item The effectiveness of feature selection methods remains relatively consistent across different backbone models. This is because the information contained in feature combinations is objective and factual and is not affected by the backbone model.
    \item The validity of feature selection methods is distinct across different datasets. The shallow feature selection methods perform better in Criteo and Aliccp. The possible reason is the feature value distribution in the other two datasets is very imbalanced, making shallow models prone to overfitting. Gate-based feature selection methods are good at dealing with limited data, while sensitivity-based feature selection methods achieve the best performance among all methods in datasets with rich samples. This finding can be attributed to the fact that in the data-scarce scenario, gradients are too sensitive to judge feature importance. However, with sufficient data support, the stability of gradient information will be greatly improved, allowing for more accurate feature importance rankings.
    \item The two-stage approach yields more stable results compared to the single-stage method across different backbone models. This is because the feature combinations filtered out by the two-stage method can exist independently after the search phase, whereas the single-stage method requires retraining the feature importance scores with each training session.
\end{itemize}

\subsection{Stability Evaluation (RQ2)} \label{K_results}
In this section, we conduct experiments on each backbone model with different number of selected features $k$ to investigate their robustness and stability. Specifically, we iterate a specific list for $k$ (e.g., $[5,10,15,17,19,21]$ for Avazu) and record the corresponding performance of feature selection methods. Then, the stability of feature selection methods can be revealed from two perspectives, visual patterns of the performance curve (Figure~\ref{fig:k}) and AUKC values calculated as Equation~\eqref{eq:AUKCCAl} (Table~\ref{overall_results_AUKC}).

In Figrue~\ref{fig:k}, the x-axis represents the number of selected features $k$ and the y-axis represents the evaluation metrics AUC or Logloss. We only visualize the results of DeepFM, for we find that the trends for different backbone models remain consistent. The detailed experimental results for the other backbone models are listed in our released repository. In addition, it is worth noting that to make OptFS applicable in this experiment, we assign importance scores to features based on the drop rates of feature values in each feature field. From Figure~\ref{fig:k}, we can find:%
\begin{itemize}[leftmargin=*]
    \item The effectiveness of features selected by different methods varies in ranking when the number of selections changes. Specifically, when the number of features selected is smaller (i.e., 5 or 10), Autofield, SHARK, and SFS perform better. They obtain information through the values or gradients of gate vectors, guiding the sorting of feature importance. They are good at selecting the most informative features from the candidate feature sets. OptFS doesn't initially perform well, as it is inherently a single-stage method. Modifying it to a two-stage method based on different feature value drop rates may not align with the goal of enhancing effectiveness through feature selection. When the number of features approaches the full set of features, the Permutation method performs well. This is because the permutation method calculates feature importance scores based on the loss of effect caused by dropping a feature from the full set. It tends to identify redundant features with less information.
    \item The XGBoost and GBDT methods show different trends on the Avazu and Criteo datasets. Specifically, they perform poorly on the Avazu dataset but well on the Criteo dataset. This is because the Avazu dataset's feature values are concentrated in a few feature fields, leading to an imbalance that makes XGBoost and GBDT prone to overfitting, which affects the assessment of feature importance~\cite{strobl2007bias}.
\end{itemize}

To offer a more direct comparison of the stability, we evaluate the selection results using the AUKC metric. Table~\ref{overall_results_AUKC} shows AUKC results on Avazu and Criteo. We can conclude: %
\begin{itemize} [leftmargin=*]
    \item Overall, sensitivity-based feature selection methods outperform gate-based feature selection methods, both of which are superior to shallow feature selection methods. The possible reason is that gradient-based methods calculate feature importance based on partial batches, which helps to prevent overfitting. The inferior performance of shallow methods may be attributed to their excessive simplicity, which cannot effectively capture the complex relationships between features to determine feature importance.
    \item Regarding specific methods, AutoField, SHARK, and SFS exhibit the best performance, demonstrating leading effects across different datasets. It is noteworthy that AutoField significantly surpasses other gate-based feature selection methods. This may be attributed to AutoField being trained in a bi-level optimization manner, which reduces the risk of overfitting, making the selection results more stable and reliable. 
\end{itemize}

\begin{table}[t]
\centering
\caption{Experimental results with performance limitations.}
\label{performance_limitation}
\resizebox{0.85\linewidth}{!}{
\begin{tabular}{ccccc} 
\toprule
\multirow{2}{*}{FS Methods} & \multicolumn{2}{c}{Avazu} & \multicolumn{2}{c}{Criteo}  \\
                            & memory remain & $k$         & memory remain & $k$           \\ 
\midrule
Lasso                       & 99.99963\%    & 17        & 68.73103\%    & 25          \\
GBDT                        & 100.00000\%   & 23        & 23.22062\%    & 30          \\
RF                          & 99.99161\%    & 10        & 99.56109\%    & 20          \\
XGBoost                     & 100.00000\%   & 23        & 75.20371\%    & 33          \\
AutoField                   & 99.91556\%    & 10        & 8.17985\%     & 20          \\
AdaFS                       & 100.00000\%   & 23        & 100.00000\%   & 39          \\
OptFS                       & 99.73494\%    & 17        & 79.05157\%    & 25          \\
LPFS                        & 99.96204\%    & 15        & 0.00100\%     & 15          \\
Permutation                 & 71.56409\%    & 15        & 48.77971\%    & 25          \\
SHARK                       & 99.98986\%    & 10        & 7.31463\%     & 15          \\
SFS                         & 99.98345\%    & 10        & 8.16199\%     & 15          \\
\bottomrule
\end{tabular}}
\end{table}

\subsection{Efficiency Analysis (RQ3)}

In this section, we answer the RQ3 from two aspects: experiments with performance limitations and with memory limitations.

\subsubsection{Experiments with performance limitations}
In real-world scenarios, an important consideration is reducing the memory usage of a model as much as possible while ensuring its effectiveness. To test the memory-saving capabilities of different methods under the constraint of maintaining effectiveness, experiments were conducted on two classic DRS datasets, Avazu and Criteo, using DCN as the backbone model. Firstly, we calculate an acceptable threshold for effectiveness based on the baseline performance of the backbone model (without feature selection, using the full set of features), considering a 1\% loss in AUC as acceptable. The threshold is AUC 0.77873 for Avazu and AUC 0.79245 for Criteo. Based on the feature importance ranking list obtained in the search phase by different feature selection methods, we select features from highest to lowest importance until the model's performance exceeds the threshold. We record the number of features and feature values required to reach this lower threshold of effectiveness for each feature selection method. Since most parameters in DRS models are concentrated in the embedding table, we only considered the memory usage of the embedding table when calculating the memory models required. From Table~\ref{performance_limitation}, we can conclude that: 
\begin{itemize}[leftmargin=*]
    \item From the perspective of memory usage, methods like AutoField, LPFS, Permutation, SHARK, and SFS perform better. This indicates that even in DRS scenarios rich in high-dimensional sparse features, these methods do not rely solely on ID-type features (e.g., userid). Instead, they effectively select informative features. It's noteworthy that AdaFS shows 100\% memory usage on both datasets. This is because AdaFS is an instance-level selection method (where gate weights vary with each input sample), and therefore cannot save memory usage.
    \item In terms of the number of features, RF, AutoField, LPFS, SHARK, and SFS only require half or fewer of the original feature set to achieve 99\% of the backbone model's performance. On the other hand, GBDT and XGBoost need more features. A possible reason for this is that GBDT and XGBoost are prone to overfitting in high-dimensional sparse DRS datasets, which can affect the judgment of feature importance.
    \item From the perspective of the datasets, the memory-saving effectiveness of various methods is less pronounced on the Avazu dataset compared to the Criteo dataset. This is because most feature values in the Avazu dataset are concentrated in just a few feature fields, whereas in Criteo, the feature values are more evenly distributed. Therefore, when these particular feature fields are indispensable sources of information for CTR prediction, the memory-saving effect becomes less noticeable.
\end{itemize}

\subsubsection{Experiments with memory limitations}

\begin{table}[t]
\centering
\caption{Experimental results with memory limitations.}
\vspace{-3mm}
\label{memory_limitation}
\resizebox{0.85\linewidth}{!}{
\begin{tabular}{ccccccc} 
\toprule
\multirow{2}{*}{FS Methods} & \multicolumn{2}{c}{25\% memory} & \multicolumn{2}{c}{50\% memory} & \multicolumn{2}{c}{75\% memory}  \\
                            & AUC     & $k$                     & AUC     & $k$                     & AUC     & $k$                      \\ 
\midrule
Lasso                       & 0.76513 & 11                    & 0.76509 & 12                    & 0.79465 & 25                     \\
GBDT                        & 0.79981 & 32                    & 0.80003 & 33                    & 0.80008 & 34                     \\
RF                          & 0.72374 & 6                     & 0.75961 & 8                     & 0.76520 & 10                     \\
XGBoost                     & 0.76282 & 16                    & 0.77942 & 21                    & 0.78809 & 30                     \\
AutoField                   & 0.79997 & 30                    & 0.80037 & 32                    & 0.80037 & 32                     \\
OptFS                       & 0.79162 & 21                    & 0.79263 & 22                    & 0.79297 & 23                     \\
LPFS                        & 0.79892 & 25                    & 0.80015 & 30                    & 0.80031 & 32                     \\
Permutation                 & 0.79462 & 24                    & 0.79782 & 30                    & 0.79782 & 30                     \\
SHARK                       & 0.79748 & 20                    & 0.79977 & 25                    & 0.79977 & 25                     \\
SFS                         & 0.79577 & 18                    & 0.80027 & 28                    & 0.80027 & 28                     \\
\bottomrule
\end{tabular}}
\vspace{-3mm}
\end{table}

Maximizing effectiveness within limited memory is also an important application scenario for feature selection. To test the capability of different methods in feature selection under memory constraints, experiments were conducted on the Criteo dataset using DCN as the backbone model. The memory limits were categorized into three levels: 25\% memory, 50\% memory, and 75\% memory. Specifically, since the parameters in DRS are mostly concentrated in the embedding table, we used the memory usage of the full feature set's embedding table as a reference. We set thresholds at 25\%, 50\%, and 75\% of the total memory to record the optimal performance achievable under memory constraints by different feature selection methods. Since AdaFS is incapable of performing a hard selection of partial feature fields, it is not included in this experiment. Based on the results presented in Table~\ref{memory_limitation}, we can draw the following conclusion: 
\begin{itemize}[leftmargin=*]
    \item Overall, gate-based feature selection and sensitivity-based feature selection methods perform better. Even when restricted to using only 25\% of the available memory, they can achieve results close to those obtained using the full set of features. This is because these methods rely on more powerful deep learning networks that can better model feature importance and are not dependent on high-dimensional sparse ID-type features in DRS.
    \item Specifically, AutoField achieves the best results because its designed controller effectively learns the importance of each field for the prediction outcome. The DARTS (Differentiable Architecture Search)~\cite{liu2018darts} parameter updating manner made the learning of the gate vector more robust and capable of generalization. In contrast, Random Forest (RF) performs the worst. This is because, in the context of high-dimensional sparse DRS, RF tends to assign high weights to ID-type features with numerous feature values. Such feature fields offer more opportunities for splitting, which facilitates the training of RF. However, this approach increases the risk of model overfitting. This tendency of RF to favor features with many splitting points can lead to a bias towards complex, less generalizable models that don't necessarily capture the most predictive or relevant features for the task at hand.
\end{itemize}
\begin{table}
\centering
\caption{Overall experimental results on industrial dataset}
\label{tab:industry}
\resizebox{\linewidth}{!}{
\begin{tabular}{ccccccc} 
\toprule
Methods & Base     & AutoField         & LPFS     & Permutation      & SHARK             & SFS       \\ 
\midrule
AUC     & 0.933920 & 0.933915          & 0.933858 & \uline{0.933946} & \textbf{0.933985} & 0.933894  \\
AUKC    & N/A      & \textbf{0.819053} & 0.817445 & \uline{0.818387} & 0.818169          & 0.817084  \\
\bottomrule
\end{tabular}}
\end{table}

\begin{figure}[t]
    \centering
    \includegraphics[width=0.8\linewidth]{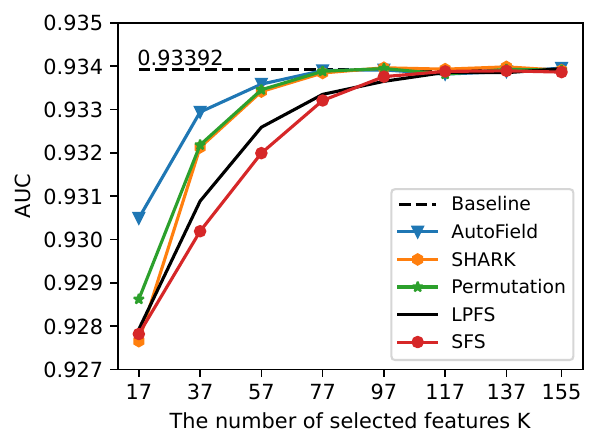}
    \vspace{-1mm}
    \caption{K-performance figure on industrial dataset}
    \label{fig:industrial_results}
    \vspace{-1mm}
\end{figure}
\subsection{Offline Results on Industrial Dataset (RQ4)}

To find whether our conclusions for feature selection methods align between large-scale industrial datasets and public datasets, we conduct comparison experiments on large-scale industrial datasets.

\subsubsection{Industrial Dataset}
The dataset is constructed with sampled CVR records from 2023-11-29 to 2023-11-30 containing approximately seven hundred million samples. 
The feature used in this offline dataset is aligned with the online service model, which contains over 157 features. 
All of these features have been roughly proved useful in offline leave-one-out experiments.

\subsubsection{Experiment} 
Considering the industry dataset is relatively huge, we only select the top 5 elite methods for comparison according to their performance on public datasets, including two gate-based methods (AutoField and LPFS) and three sensitivity-based methods (SHARK, Permutation, and SFS). 

The feature preprocessing is slightly different from the public dataset. Specifically, we utilize AutoDis \cite{autodis} technique to directly convert dense features into embeddings with the same size as sparse features. As for multi-hot features, we adopt the reduce-mean operator to aggregate multiple embeddings into one feature embedding. Finally, the preprocessed features are fed into the FibiNet backbone.

After training, we grab each method's feature importance and perform top-k experiments. The experiment result is demonstrated in Table~\ref{tab:industry} and Figure \ref{fig:industrial_results}. From the redundant feature elimination perspective, SHARK, AutoField, and Permutation perform the best. They eliminate about half of the feature sets (80 features) without decreasing model accuracy. 
However, from an effective feature mining perspective, AutoField performs best for small feature sets, aligning with our observations from the public datasets. We can also find that AutoField achieves the highest AUKC in Table~\ref{tab:industry}, demonstrating its robustness and stability across different numbers of selections. LPFS's performance is slightly misaligned with the one on the public dataset. We attribute this misalignment to the sensitive learnable polarization module.

\subsection{Online Experiments (RQ5)}
To validate the effectiveness of our benchmark in online environments, we reconfigure partial benchmarks to adapt to our feature factory, package them into a new toolkit, and conduct online tasks with this toolkit for redundant feature elimination tasks. 

Specifically, we create an experiment group that reduces 30\% features with our toolkit in a latency-sensitive online recommendation platform.
Then, we deploy the experimental group for the A/B test with 5\% traffic (approximately 1 million users). After one week's observation, the inference latency was reduced by approximately 20\% while the number of video views (vv) and average play duration remained the same. This experiment proves the effectiveness of our benchmark toolkits in the online environment.

\subsection{Similarity Visualization (RQ6)} \label{similarity}

\begin{figure}[t]
    \centering
    \includegraphics[width=0.9\linewidth]{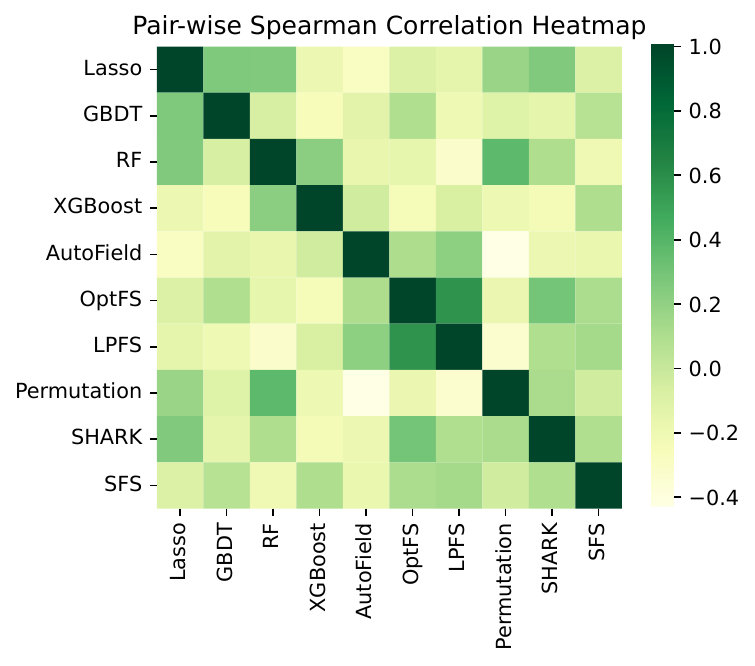}
    \caption{Similarity between feature selection methods.}
    \label{fig:spearman}
\end{figure}
In this section, we visualize the similarity between feature rankings of different feature selection methods in DRS across 4 public datasets. Figure~\ref{fig:spearman} shows the heatmap of pair-wise Spearman correlation similarities. In the heatmap, the x-axis and y-axis represent different feature selection methods. The color of each square indicates the level of similarity. The more the color leans towards dark green, the more similar between the feature rankings derived from the two feature selection methods. Conversely, the more the color leans towards light green, the greater the difference in the feature rankings selected by the two methods.

We can discern the following information from the heatmap: 1) The heatmap displays three darker clusters in the upper left, middle, and lower right. This is because the features selected by the three different category methods are more similar within each category. 2) The rankings of LPFS and OptFS are highly similar. This is because they are both gate-based methods and use the L0 regularization term to control the sparsity of the gates. 3) The rankings of AutoField and Permutation are the most dissimilar. This aligns with our conclusion in Section~\ref{K_results}. AutoField, as a gate-based method, determines feature importance by learning the gate weights of different feature fields during the model's training process, aiding in uncovering features rich in information. On the other hand, Permutation, as a sensitivity-based feature selection method, judges feature importance based on the loss of effect caused by shuffling a particular feature. This approach is more conducive to identifying features with less information. Therefore, their rankings show significant differences. 4) The rankings of SHARK and SFS are quite similar because they both determine feature importance based on the gradient of the features, sharing the same source of information.
\section{Related Works}

\noindent\textbf{Feature Selection Methods.}
Machine learning has shown superior results in data mining~\cite{jia2024g3,zhang2024m3oe,li2023agent4ranking,jia2024d3,karpathy2015deep,jia2023mill,jia2024fine}. The modeling performance relies heavily on input features, making feature selection a crucial part of feature engineering.
Traditional feature selection methods fall into three categories: filter, wrapper, and embedded methods~\cite{guyon2003introduction}. Filter methods utilize criteria to identify predictive feature fields, exemplified by the Chi2 score~\cite{liu1995chi2} and mutual information~\cite{vergara2014review}. Wrapper methods employ black-box models to select predictive features, assessing the utility of feature subsets through, often, generic algorithms~\cite{granitto2006recursive,shah2004data}. Embedded methods, on the other hand, integrate the feature selection process within the prediction model, thereby evaluating the effectiveness of feature subsets in conjunction. Notable embedded methods include LASSO~\cite{tibshirani1996lasso} and Gradient Boosting Machine~\cite{friedman2001greedy}. The advent of deep learning has spurred novel approaches~\cite{liu2024multifs,chen2024comprehensive,zhaok2021autoemb,zhao2021autodim,liu2020automated,zhao2021autoloss,li2023automlp}, such as gate-based methods, where learnable gates determine the significance of feature fields~\cite{wang2022autofield,lin2022adafs,lyu2023optfs,lee2023mvfs}, and sensitivity-based methods, leveraging gradient techniques to identify critical features by their sensitivity~\cite{Fperm,wang2023sfs,fisher2019allpermutation}. Furthermore, some studies explored reinforcement learning to automate feature selection, using agents to pinpoint predictive features~\cite{liu2019automating,fan2020autofs,zhao2020simplifying}.

Considering their appliability to DRS, our benchmark focus on traditional embedded methods (termed ``shallow methods''), alongside gate-based and sensitivity-based methods. Other techniques often fall short in terms of effectiveness or efficiency within the DRS context. Specifically, filter methods might overlook the DRS model's nuances by relying solely on data-intrinsic relationships, leading to suboptimal outcomes. Besides, wrapper and reinforcement learning methods, typically designed for smaller datasets, become impractically time-consuming for the expansive datasets characteristic of DRS, which can encompass millions of samples.

\noindent\textbf{Benchmarks for Feature Selection.} 
Most existing benchmarks are tailored for classical downstream models~\cite{bolon2014review,bommert2020benchmark} and rely on synthetic or domain-specific datasets~\cite{bolon2013review,forman2003extensive,darshan2018performance}. Due to their constrained relevance, these benchmarks frequently fall short in providing the necessary guidance to propel research forward in the domain of DRS feature selection. The most related work, DeepLasso~\cite{deeplasso}, confines its evaluation to shallow methods and a narrow set of backbone models, primarily targeting tabular learning tasks. Despite utilizing real-world datasets, the scale of these datasets pales in comparison to those encountered in DRS scenarios, thus limiting the provision of actionable insights for DRS feature selection strategies. ERASE undertakes a thorough evaluation of shallow, gate-based, and sensitivity-based methods across an array of representative DRS backbones, leveraging both large-scale public datasets and private industrial datasets. Furthermore, ERASE proposes a novel metric designed to assess feature selection efficacy comprehensively.

\section{Conclusion}
In this study, we introduce \textbf{ERASE}, a comprehensive b\textbf{E}nchma\textbf{R}k for fe\textbf{A}ture \textbf{SE}lection for deep recommender systems (DRS). ERASE integrates a broad spectrum of feature selection methods pertinent to DRS and ensures fair and comprehensive experimentation across four public datasets as well as real-world industrial datasets. Furthermore, ERASE pioneers the adoption of the AUKC metric, devised to address the shortcomings of existing metrics by offering a thorough evaluation of the robustness and stability of feature selection methods over various numbers of selected features. The empirical validation provided by both offline and online analyses on industrial datasets underscores the applicability of our findings in practical settings, offering valuable perspectives on the deployment of feature selection methods within DRS environments.

\section{Acknowledgments}
This research was partially supported by Huawei (Huawei Innovation Research Program), Research Impact Fund (No.R1015-23), APRC - CityU New Research Initiatives (No.9610565, Start-up Grant for New Faculty of CityU), CityU - HKIDS Early Career Research Grant (No.9360163), Hong Kong ITC Innovation and Technology Fund Midstream Research Programme for Universities Project (No.
ITS/034/22MS), Hong Kong Environmental and Conservation Fund (No. 88/2022), and SIRG - CityU Strategic Interdisciplinary Research Grant (No.7020046, No.7020074).

\clearpage
\balance
\bibliographystyle{ACM-Reference-Format}
\bibliography{reference}

\appendix
\begin{table}[ht]
\centering
\caption{Dataset statistics}
\label{dataset}
\resizebox{\linewidth}{!}{
\label{table:dataset_statistics}
\begin{tabular}{ccccc} 
\toprule
Dataset          & Avazu      & Criteo     & Movielens-1M & AliCCP      \\ 
\midrule
Interactions     & 40,428,967 & 45,850,617 & 1,000,209    & 85,316,519  \\
Users            & N/A          & N/A          & 6,040        & 238,635     \\
Items            & N/A          & N/A          & 3,706        & 467,298     \\
Interaction Type & Click      & Click      & Rating (1-5)       & Click       \\
Feature Num      & 23         & 39         & 9            & 23          \\
\bottomrule
\end{tabular}}
\end{table}
\section{Datasets} \label{appendix:datasets}

In this subsection, we introduce the datasets utilized in our benchmark. The statistics of these datasets are listed in Table~\ref{table:dataset_statistics}. We select the following four datasets for use:

\begin{itemize}[leftmargin=*]
    \item \textbf{Avazu.} Avazu is a commonly used dataset in the field of feature selection for deep recommendation systems. It consists of 23 features and 40,428,967 interaction records. We divide the dataset into training, validation, and test sets in a 7:2:1 ratio.
    \item \textbf{Criteo.} Criteo is another frequently used dataset for studying feature selection in Deep Recommendation Systems. The Criteo dataset contains 45,850,617 samples and 39 features, offering a larger number of features. We also adopt a 7:2:1 ratio for dividing the data into training, validation, and test sets to train the model.
    \item \textbf{Movielens-1m.} The Movielens dataset is a well-known public movie dataset in the field of recommendation systems, featuring multiple variants with different data volumes to meet diverse research needs. We choose the Movielens-1M dataset, which comprises 9 features. The limited number of features presents a greater challenge for feature selection methods. In our experiments with Movielens, we also divide the dataset into training, validation, and test sets according to a 7:2:1 ratio. The interactions with a rating bigger than 3 are considered as positive samples.
    \item \textbf{AliCCP.} Alibaba Click and Conversion Prediction (AliCCP) dataset is extracted from the real-world e-commerce platform Taobao. It is a popular dataset used for Click-Through Rate (CTR) estimation. It consists of 23 features and 85,316,519 samples, making it an effective tool for evaluating feature selection methods in scenarios closely resembling real advertising contexts. We follow the original AliCCP splitting approach~\cite{ma2018esmm}, allocating 50\% of the data for training, with the remaining data split equally into validation and test sets in a 1:1 ratio.
\end{itemize}

\section{Backbone Models} \label{appendix:bbm}

we apply a variety of feature selection methods to the following four popular deep recommendation models in this work.

\begin{itemize}[leftmargin=*]
    \item \textbf{Wide\&Deep~\cite{cheng2016wide}.} Wide and Deep (Wide\&Deep) model is developed by Google to improve the recommender systems. It combines wide and deep models to fit specific feature combinations in large-scale datasets and previously unseen feature combinations through low-dimensional dense embeddings.
    \item \textbf{DeepFM~\cite{guo2017deepfm}.} DeepFM is an effective model in recommender systems. It combines the strengths of FM models and deep neural networks to extract both low-order feature interactions and high-order feature interactions.
    \item \textbf{DCN~\cite{wang2017deep}.} The Deep \& Cross Network (DCN) model is proposed by Google to capture both explicit and implicit feature interactions for prediction tasks. The cross layers have cross operations that learn complex feature interactions to improve the prediction performance.
    \item \textbf{FibiNet~\cite{huang2019fibinet}.} The Feature Importance and Bilinear feature Interaction NETwork (FibiNet) is an innovative model in CTR prediction. It introduces the SENET and bilinear layer. SENET learns feature importance adaptively and the bilinear layer captures high-order feature interactions.
\end{itemize}

\begin{table*}[h]
\centering
\caption{Overall experimental results of feature selection for deep recommender systems.}
\label{appendix_overall_results}
\resizebox{0.7\textwidth}{!}{
\begin{tabular}{cccccccccc} 
\toprule
\multirow{2}{*}{Backbone model} & \multirow{2}{*}{Methods} & \multicolumn{2}{c}{Avazu}           & \multicolumn{2}{c}{Criteo}          & \multicolumn{2}{c}{Movielens-1M}    & \multicolumn{2}{c}{AliCCP}           \\ 
\cmidrule[\heavyrulewidth]{3-10}
                                &                          & AUC              & Logloss          & AUC              & Logloss          & AUC              & Logloss          & AUC              & Logloss           \\ 
\toprule
\multirow{12}{*}{DeepFM}        & no\_selection            & 0.78576          & 0.37680          & 0.80024          & \textbf{0.45321} & 0.79027          & 0.54124          & 0.61813          & 0.16176           \\
                                & Lasso                    & 0.78587          & 0.37654          & 0.79832          & 0.45483          & \uline{0.80683}  & \uline{0.52498}  & 0.61887          & 0.16174           \\
                                & GBDT                     & 0.76930          & 0.38571          & 0.80011          & 0.45329          & 0.78942          & 0.54234          & 0.61864          & \textbf{0.16150}  \\
                                & RF                       & \uline{0.78642}  & 0.37630          & 0.79920          & 0.45427          & 0.78920          & 0.54217          & 0.61772          & 0.16172           \\
                                & XGBoost                  & 0.76953          & 0.38548          & 0.80022          & 0.45333          & 0.78974          & 0.54190          & 0.61822          & 0.16172           \\
                                & AutoField                & 0.78611          & 0.37641          & 0.80022          & \uline{0.45325}  & \textbf{0.80722} & \textbf{0.52471} & \uline{0.61894}  & 0.16174           \\
                                & AdaFS                    & 0.78278          & 0.37922          & 0.79950          & 0.45405          & 0.78590          & 0.54676          & 0.61726          & \uline{0.16158}   \\
                                & OptFS                    & 0.78624          & \uline{0.37627}  & 0.79934          & 0.45505          & 0.79027          & 0.54124          & 0.61551          & 0.16195           \\
                                & LPFS                     & \textbf{0.78840} & \textbf{0.37601} & \textbf{0.80080} & 0.45364          & 0.78885          & 0.54276          & \textbf{0.61941} & 0.17864           \\
                                & Permutation              & 0.78624          & 0.37640          & 0.80006          & 0.45349          & 0.78974          & 0.54207          & 0.61844          & 0.16174           \\
                                & SHARK                    & 0.78611          & 0.37643          & \uline{0.80035}  & 0.45329          & 0.78888          & 0.54342          & 0.61840          & 0.16174           \\
                                & SFS                      & 0.78626          & 0.37631          & 0.80020          & 0.45330          & 0.78982          & 0.54194          & 0.61835          & 0.16168           \\ 
\midrule
\multirow{12}{*}{Wide\&Deep}      & no\_selection            & 0.78574          & 0.37679          & 0.79971          & 0.45346          & 0.79041          & 0.54163          & 0.62137          & 0.16154           \\
                                & Lasso                    & 0.78578          & 0.37665          & 0.79761          & 0.45526          & \textbf{0.80750} & \textbf{0.52400} & 0.62096          & 0.16159           \\
                                & GBDT                     & 0.76900          & 0.38583          & 0.79979          & 0.45337          & 0.78926          & 0.54252          & \uline{0.62187}  & 0.16146           \\
                                & RF                       & 0.78571          & 0.37664          & 0.79907          & 0.45400          & 0.78992          & 0.54170          & \textbf{0.62202} & 0.16132           \\
                                & XGBoost                  & 0.76886          & 0.38587          & \uline{0.80023}  & 0.45320          & 0.78893          & 0.54266          & 0.62174          & 0.16142           \\
                                & AutoField                & 0.78601          & 0.37653          & 0.79971          & 0.45348          & \uline{0.80736}  & \uline{0.52496}  & 0.62176          & 0.16134           \\
                                & AdaFS                    & \textbf{0.78607} & 0.37665          & 0.79952          & 0.45394          & 0.78703          & 0.54467          & 0.62106          & \uline{0.16123}   \\
                                & OptFS                    & 0.78582          & 0.37654          & \textbf{0.80111} & \textbf{0.45284} & 0.79041          & 0.54163          & 0.62053          & 0.16144           \\
                                & LPFS                     & 0.78606          & 0.37680          & 0.79953          & 0.45477          & 0.79072          & 0.54136          & 0.61916          & 0.17177           \\
                                & Permutation              & \textbf{0.78607} & \uline{0.37647}  & 0.79988          & 0.45342          & 0.78936          & 0.54226          & 0.62156          & 0.16138           \\
                                & SHARK                    & 0.78592          & 0.37655          & 0.80019          & \uline{0.45313}  & 0.78980          & 0.54146          & 0.62148          & 0.16154           \\
                                & SFS                      & 0.78592          & \textbf{0.37646} & 0.80010          & 0.45326          & 0.78993          & 0.54139          & 0.62165          & \textbf{0.16122}  \\ 
\midrule
\multirow{12}{*}{DCN}           & no\_selection            & 0.78660          & 0.37615          & 0.80045          & 0.45297          & 0.78966          & 0.54228          & \uline{0.62315}          & 0.16134           \\
                                & Lasso                    & 0.78628          & 0.37638          & 0.79860          & 0.45456          & \uline{0.80830}  & \uline{0.52367}  & 0.62289          & 0.16124           \\
                                & GBDT                     & 0.76926          & 0.38578          & 0.80049          & 0.45299          & 0.79106          & 0.54048          & 0.62318          & 0.16131           \\
                                & RF                       & 0.78644          & 0.37614          & 0.79978          & 0.45342          & 0.79058          & 0.54058          & 0.62310          & 0.16156           \\
                                & XGBoost                  & 0.76934          & 0.38565          & \uline{0.80067}  & \uline{0.45264}  & 0.79102          & 0.54050          & \textbf{0.62321}  & 0.16131           \\
                                & AutoField                & 0.78663          & 0.37620          & 0.80051          & 0.45286          & \textbf{0.80932} & \textbf{0.52226} & 0.62294 & \textbf{0.16114}  \\
                                & AdaFS                    & \textbf{0.78706} & \textbf{0.37597} & 0.80018          & 0.45336          & 0.78792          & 0.54366          & 0.62264          & 0.16122           \\
                                & OptFS                    & 0.78663          & 0.37618          & \textbf{0.80164} & \textbf{0.45222} & 0.78966          & 0.54228          & 0.62083          & 0.16183           \\
                                & LPFS                     & \uline{0.78686}  & 0.37631          & 0.80048          & 0.45349          & 0.79006          & 0.54123          & 0.62202          & 0.16536           \\
                                & Permutation              & 0.78651          & 0.37616          & 0.80052          & 0.45336          & 0.79047          & 0.54086          & 0.62311          & \uline{0.16116}   \\
                                & SHARK                    & 0.78656          & \uline{0.37614}  & 0.80063          & 0.45300          & 0.79116          & 0.54019          & 0.62302          & 0.16127           \\
                                & SFS                      & 0.78649          & 0.37620          & 0.80066          & 0.45273          & 0.79066          & 0.54067          & 0.62299          & 0.16118           \\ 
\midrule
\multirow{12}{*}{FibiNet}       & no\_selection            & 0.79039          & 0.37401          & 0.80528          & 0.44854          & 0.78490          & 1.44348          & 0.62138          & 0.16138           \\
                                & Lasso                    & 0.79057          & 0.37399          & 0.80345          & 0.45012          & \textbf{0.80350} & 0.56017          & 0.62147          & 0.16162           \\
                                & GBDT                     & 0.77222          & 0.38403          & 0.80537          & 0.44846          & 0.78213          & 1.44155          & \uline{0.62257}  & 0.16145           \\
                                & RF                       & 0.79053          & 0.37377          & 0.80451          & 0.44917          & 0.78421          & 0.61107          & 0.62166          & \uline{0.16133}   \\
                                & XGBoost                  & 0.77202          & 0.38436          & 0.80554          & \uline{0.44828}  & 0.78356          & 1.25796          & 0.62132          & 0.16147           \\
                                & AutoField                & 0.79068          & 0.37371          & 0.80523          & 0.44864          & \uline{0.80091}  & \uline{0.54683}  & \textbf{0.62272} & 0.16137           \\
                                & AdaFS                    & 0.78959          & 0.37441          & 0.80439          & 0.44953          & 0.78478          & 0.54911          & 0.61938          & 0.16158           \\
                                & OptFS                    & 0.79059          & 0.37383          & 0.80246          & 0.45159          & 0.78490          & 1.44348          & 0.61958          & 0.16154           \\
                                & LPFS                     & \textbf{0.79093} & 0.37385          & \textbf{0.80559} & 0.44898          & 0.78790          & \textbf{0.54625} & 0.61750          & 0.16797           \\
                                & Permutation              & 0.79071          & 0.37405          & 0.80515          & 0.44861          & 0.78457          & 0.67354          & 0.62136          & 0.16161           \\
                                & SHARK                    & \uline{0.79078}  & \textbf{0.37364} & 0.80550          & 0.44833          & 0.78574          & 1.33867          & 0.62198          & \textbf{0.16123}  \\
                                & SFS                      & 0.79068          & \uline{0.37369}  & \uline{0.80557}  & \textbf{0.44826} & 0.78507          & 1.20824          & 0.62219          & 0.16140           \\
\bottomrule
\end{tabular}}
\end{table*}

\section{Metrics} \label{appendix:metrics}

In this section, we detail the two frequently used metrics in recommender systems: AUC and Logloss.

1) \textbf{AUC.} Area Under the ROC Curve (AUC) measures the area under the ROC curve. The ROC curve is one graph shows that the classification performance of the model under different thresholds. In addition, in random positive-negative pairs, AUC also represents the probability that the model ranks the positive sample with a higher score than the negative one. It can be formalized as follows:
\begin{equation}
    AUC = \frac{\sum_{i}^{M}rank_i - M(1+M)/2}{M \times N}, \ i \in positiveClass
\end{equation}
where $M$ denotes the number of positive samples and $N$ is the number of negative samples.

2) \textbf{Logloss.} Logloss is also named Binary Cross-Entropy Loss, it is always used in binary classification tasks. The formula for Logloss is described as follows:
\begin{equation}
    Logloss = -\frac{1}{N} \sum_{i=1}^{N} [y_i log(\hat{y}_i) + (1-y_i)log(1-\hat{y_i})]
\end{equation}
where $N$ is the number of samples, $y_i$ denotes the predicted label for $i$th sample, and $\hat{y}_i$ denotes the ground truth for $i$th sample.

\section{Formula Derivation of AUKC} \label{AUKC}

In this section, we detail the formula derivation process of AUKC. AUKC is a comprehensive metric for assessing the AUC performance of a given feature selection method across different numbers of feature selections. Under ideal circumstances, experiments can be conducted for each number of selections to obtain the corresponding AUC metrics, and then derive AUKC:
\begin{align}
    AUKC &= \frac{\sum_{k=1}^{|K|}((AUC_k-0.5)+(AUC_{k+1}-0.5)) / 2}{|K|/2} \\
    &= \frac{1}{|K|}\sum_{k=1}^{|K|}(AUC_k+AUC_{k+1}-1)
\end{align}
where $|K|$ denotes the number of all features, $AUC_k$ is the AUC metric with $k$ features selected. When there is no feature selected, the model will predict labels randomly (i.e., $AUC=0.5$). Therefore, we subtract 0.5 from $AUC_{k}$ to obtain the improvement. 

When the number of selected features is unevenly distributed across the length of the set of features (e.g., in practice, the granularity of the feature selection number often becomes finer as it approaches the total number of features, and coarser when there are fewer features), AUKC has a more general format:
\begin{align}
    AUKC &= \frac{\sum_{n=1}^{|N|}((AUC_{nl}-0.5)+(AUC_{nr}-0.5)) \times \Delta l_{n} / 2}{|K|/2} \\
    &= \frac{1}{|K|}\sum_{n=1}^{|N|}(AUC_{n,l}+AUC_{n,r}-1) \times \Delta l_{n}
\end{align}
where $|N|$ represents the number of segments for the number of features selected across the entire length of the feature set. For example, with 6 features, if experiments are conducted at feature counts of 2,4,5, and 6, then $|N|=4$. $AUC_{n,l}$ is the AUC corresponding to the number of features selected at the left endpoint of the $n$th segment, and $AUC_{n,r}$ denotes the AUC with the number of selections at the right endpoint of the $n$th segment. $\Delta l_{n}$ represents the length of the $n$th segment.

\section{Complete Overall Performance} \label{appendix:overall}

Due to page limitations, we can not include the complete overall experimental results in the main content. We list the complete results in Table~\ref{appendix_overall_results} for reference.

\end{document}